\pdfoutput=1

\documentclass[11pt]{article}

\usepackage{xurl}
\usepackage[breaklinks=true]{hyperref}

\usepackage[final]{acl}

\usepackage{times}
\usepackage{latexsym}
\usepackage{multicol}
\usepackage{latexsym}
\usepackage{graphicx}
\usepackage{amsmath}
\usepackage{expex}
\usepackage{float}
\usepackage{changepage}
\usepackage{textcomp}
\usepackage[title]{appendix}

\usepackage{pifont}

\usepackage{booktabs}
\usepackage{makecell}
\newcommand{\tabincell}[2]{\begin{tabular}{@{}#1@{}}#2\end{tabular}}
\usepackage{multirow}
\PassOptionsToPackage{table}{xcolor}

\usepackage[T1]{fontenc}

\usepackage[utf8]{inputenc}

\usepackage{microtype}

\usepackage{inconsolata}

\usepackage{graphicx}

\usepackage{xspace}  
\usepackage{tipa} 

\usepackage{url}      

\sbox0{T} 
\sbox1{\rule[0.5\ht0]{0.6\wd0}{0.5pt}} 

\newcommand\textTstroketitle{%
  \makebox[\dimexpr\wd0/2-\wd1/2\relax][l]{T}%
  \rule[0.6\ht0]{1\wd0}{0.8pt}%
}

\newcommand\textTstrokesectitle{%
  \makebox[\dimexpr\wd0/2-\wd1/2\relax][l]{T}%
  \rule[0.45\ht0]{1\wd0}{0.8pt}%
}

\newcommand\textTstroke{%
  \makebox[\dimexpr\wd0/2-\wd1/2\relax][l]{T}%
  \rule[0.4\ht0]{0.55\wd0}{0.5pt}%
}

\newcommand{\sencotentitle}{SEN\'{C}O\textTstroketitle \hspace*{0.15em}EN\xspace}

\newcommand{\sencotensectitle}{SEN\'{C}O\textTstrokesectitle \hspace*{0.15em}EN\xspace}

\newcommand{\sencoten}{SEN\'{C}O\textTstroke \hspace*{0.15em}EN\xspace}

\newcommand{\wsanec}{\textsubbar{W}S\'{A}NE\'{C}\xspace}

\newcommand{\cedilla}{\hspace{0.2em}\c{}\hspace{0.2em}}

\newcommand{\xmark}{\ding{55}}
\newcommand{\cmark}{\ding{51}}

\title{Supporting \sencotentitle Language Documentation Efforts \\ with Automatic Speech Recognition}

\author{
    Mengzhe Geng$^{1*}$~
    Patrick Littell$^1$~
    Aidan Pine$^1$~
    PENÁĆ$^2$~
    Marc Tessier$^1$~
    Roland Kuhn$^1$ \\
    $^1$National Research Council Canada, Ottawa, ON, Canada \\
    \texttt{first.last@nrc-cnrc.gc.ca} \\
    $^2$\wsanec School Board, Brentwood Bay, BC, Canada \\
}

\begin{document}
\maketitle

\begin{abstract}
The \sencoten language, spoken on the Saanich peninsula of southern Vancouver Island, is in the midst of vigorous language revitalization efforts to turn the tide of language loss as a result of colonial language policies. To support these on-the-ground efforts, the community is turning to digital technology. Automatic Speech Recognition (ASR) technology holds great promise for accelerating language documentation and the creation of educational resources. However, developing ASR systems for \sencoten is challenging due to limited data and significant vocabulary variation from its polysynthetic structure and stress-driven metathesis. To address these challenges, we propose an ASR-driven documentation pipeline that leverages augmented speech data from a text-to-speech (TTS) system and cross-lingual transfer learning with Speech Foundation Models (SFMs). An n-gram language model is also incorporated via shallow fusion or n-best restoring to maximize the use of available data. Experiments on the \sencoten dataset show a word error rate (WER) of 19.34\% and a character error rate (CER) of 5.09\% on the test set with a 57.02\% out-of-vocabulary (OOV) rate. After filtering minor cedilla-related errors, WER improves to 14.32\% (26.48\% on unseen words) and CER to 3.45\%, demonstrating the potential of our ASR-driven pipeline to support \sencoten language documentation.
\end{abstract}


\let\oldthefootnote\thefootnote 
\renewcommand{\thefootnote}{}  
\footnotetext{$^*$Corresponding author.} 
\let\thefootnote\oldthefootnote 

\section{Introduction}

Language documentation often plays an important role in the revitalization of Indigenous languages. Language revitalization is, in turn, crucially important for preserving the cultural heritage and identity of Indigenous communities. \sencoten (str), the language of the \wsanec people, has faced considerable challenges, largely due to the cumulative effects of historical marginalization and cultural suppression~\citep{haque2015indigenous, LanguageRevitalizationpine_turin}. With a sharp reduction in fluent speakers, many Indigenous languages in Canada, including \sencoten, are at a critical juncture. Of the approximately 70 Indigenous languages in Canada, many urgently require revitalization efforts to prevent further loss~\citep{littell2018indigenous}. In this context, Automatic Speech Recognition (ASR) technology offers significant potential for language revitalization by supporting the transcription of spoken language, thereby potentially accelerating the development of educational curriculum developed from audio data~\citep{jimerson-prudhommeaux-2018-asr, foley18_sltu,littell2018indigenous, gupta2020automatic, gupta2020speech,liu-etal-2022-enhancing,rodriguez2023speech}. While ASR technologies have made significant strides for widely spoken languages~\citep{peddinti2015time, chan2016listen, wang2020transformer, gulati2020conformer, hu2022neural, li2023audio, hu2024self}, research on ASR systems for Canadian Indigenous languages~\citep{gupta2020automatic, gupta2020speech} remains limited.

\sencoten, also known as the Saanich language, is spoken around the Saanich peninsula in the southern region of Vancouver Island and on neighboring islands in the Strait of Georgia. The language is written with a distinct alphabet developed by the late Dave Elliott Sr.~\citep{firstvoices_sencoten}. As of 2022, there are a reported 16 fluent \sencoten speakers and 165 semi-speakers~\citep{fpcc2022status}. While ongoing and vigorous revitalization efforts~\citep{brand2002language,jim2016wsanec,bird2017role,bird2020pronunciation,elliott2024saltwater,SGILE2025} are in place, there have been no prior efforts to leverage ASR techniques to support the documentation and revitalization of \sencoten.

This paper aims to address the gap by investigating cutting-edge ASR-based techniques that can support \sencoten\ language documentation efforts. However, developing ASR systems for \sencoten\ presents two major challenges: \textbf{1) Limited data resources}: Compared with high-resource languages like English, there are very few digitized materials in \sencoten~\citep{pine-etal-2022-requirements}, and even fewer audio recordings are available with aligned transcriptions; \textbf{2) Extensive vocabulary variation}: Beyond the relatively polysynthetic nature of \sencoten, metathesis driven by stress patterns further contributes to the vast number of possible word forms as illustrated below from~\citet[Section 2.3.5.4.3]{montler1986outline}:

\begin{multicols}{2} 
    \raggedcolumns
    \lingset{glspace=!0pt plus .2em}
    \ex[glhangstyle=none]
    \let\\=\textsc
    \begingl
    \gla \textsubbar{T}QET //
    \glb \textipa{\textcrlambda}'k\textsuperscript{w}'\'{\textschwa}t//
    \glft `Put it out (a fire).'//
    \endgl
    \xe 
    
    \lingset{glspace=!0pt plus .2em}
    \ex[glhangstyle=none]
    \let\\=\textsc
    \begingl
    \gla \textsubbar{T}EQT SEN //
    \glb\textipa{\textcrlambda}'\'{\textschwa}k\textsuperscript{w}'t s{\textschwa}n//
    \glft `I’m putting it out.'//
    \endgl
    \xe 
\end{multicols}

\noindent Such morphological and phonological complexity makes it impractical to construct a sufficiently large dictionary. As a result, many words to be transcribed are absent from the system’s training data (i.e., out of vocabulary). These two challenges, taken together, significantly hinder the development of robust ASR systems for \sencoten.

To tackle the challenges associated with the development of ASR systems for \sencoten, this paper explores a range of state-of-the-art techniques, with an emphasis on end-to-end (E2E) models. E2E approaches offer a distinct advantage over traditional GMM-HMM or hybrid DNN-based systems, as they eliminate the need for a fixed lexicon. Given \sencoten's highly complex morphology, as well as the difficulty of building an exhaustive lexicon, E2E models are particularly well-suited to the task.  However, a major drawback of E2E systems is their reliance on large datasets, which poses a significant obstacle for low-resource languages like \sencoten. To address this, we propose two strategies: \textbf{1) ASR data augmentation} through a carefully designed text-to-speech (TTS) synthesis pipeline, and \textbf{2) cross-lingual transfer learning} leveraging speech foundation models (SFMs). Additionally, we incorporate an external n-gram language model (LM) using either shallow fusion~\citep{kannan2018analysis} or n-best rescoring~\citep{chow1989n} to make the most of the available data.

Experiments were conducted using the \sencoten speech dataset, comprising 4 hours of recorded audio from the ``Speech Generation for Indigenous Language Education project~\citep{SGILE2025}.  The results show that systems employing cross-lingual transfer learning with speech foundation models significantly outperformed conventional hybrid time-delay neural networks (TDNNs), particularly when recognizing unseen words not present in the training set. Moreover, incorporating TTS-synthesized data in ASR training and an external n-gram LM further enhanced system performance.

\begin{figure*}[!t]
    \centering
    \includegraphics[width=0.72\textwidth]{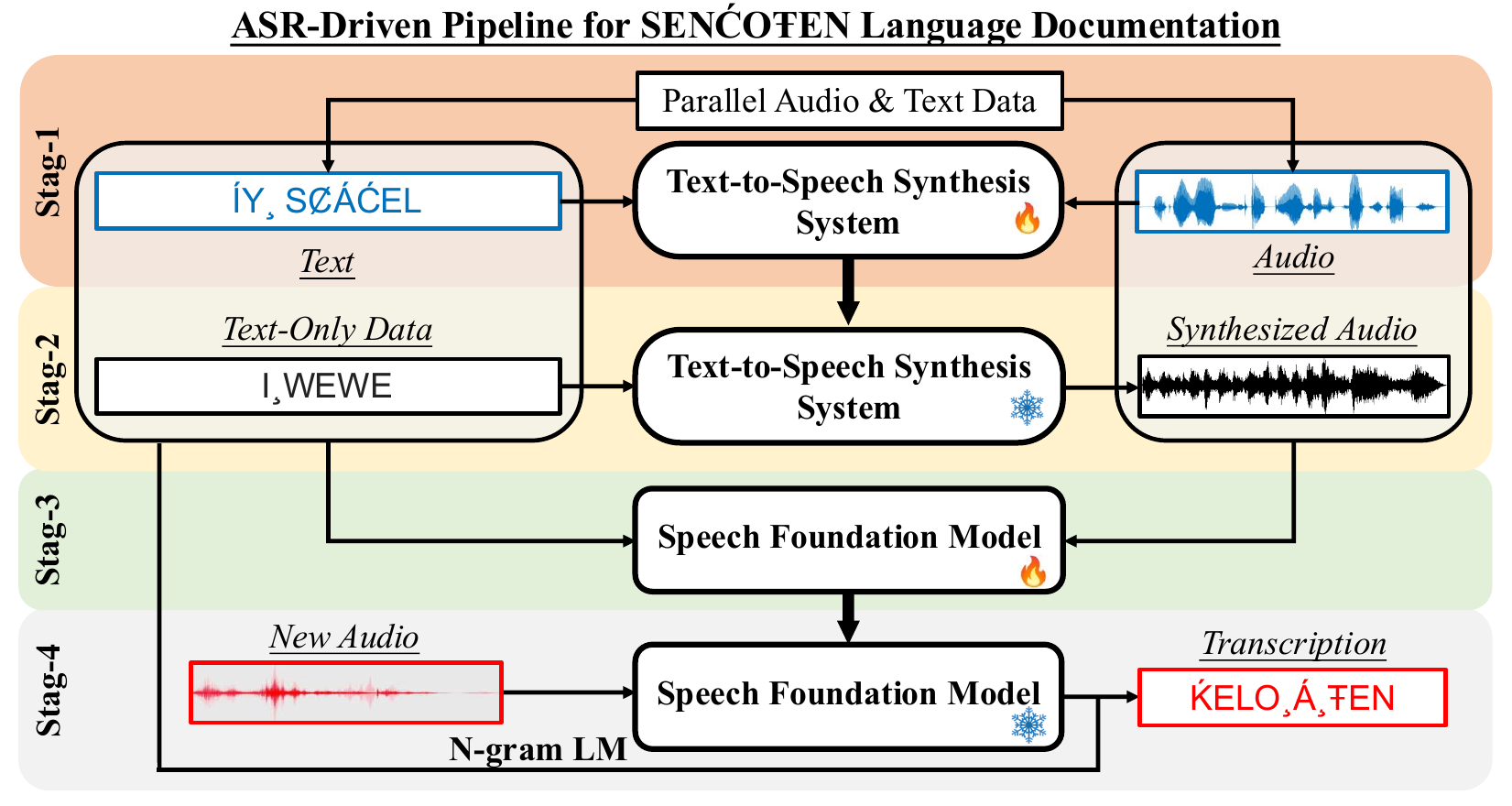}
    \caption{The proposed multi-stage ASR-driven pipeline to support \sencoten language documentation.}
    \label{fig:workflow}
\end{figure*}

This paper's key contributions are below:

\vspace{-3pt}
\begin{enumerate}
    \item \textbf{First comprehensive investigation of SFMs for documenting low-resource languages}: 
    This study represents the first systematic investigation of speech foundation models for the development of ASR systems aimed at supporting the documentation of Canadian Indigenous languages. Prior research on languages such as Inuktitut~\citep{gupta2020automatic}, Cree~\citep{gupta2020speech}, and other North American Indigenous languages, including Hupa~\citep{liu-etal-2022-enhancing}, has predominantly employed hybrid ASR architectures. These approaches typically demand in-depth linguistic expertise, particularly for the careful design and selection of subword units. While models such as Wav2vec2, XLS-R and Whisper have been explored for language documentation tasks~\citep{jimerson2023unhelpful,rodriguez2023speech}, including for languages like Hupa~\citep{jimerson2023unhelpful,venkateswaran2024looking}, Seneca~\citep{jimerson2023unhelpful} and Oneida~\citep{jimerson2023unhelpful}, our work is the first to conduct a comprehensive investigation of pre-trained SFMs in this context. By leveraging these models, we aim to widen the so-called ``transcription bottleneck'' and accelerate language documentation efforts.
    
    \item \textbf{First ASR-driven documentation pipeline for the \sencoten language}: This work introduces the first ASR-driven documentation pipeline tailored for \sencoten. To address the challenges of limited data and high lexical variation, we adopt a two-pronged strategy: ASR data augmentation via TTS and cross-lingual transfer learning based on SFMs. Moreover, we perform a systematic analysis of ASR performance under more extreme conditions, reducing the available training data to as little as 10 minutes\footnote{Details can be found in the Appendix.}.

    \item \textbf{Promising ASR performance with extended error analysis}: The top-performing system, integrating cross-lingual transfer learning, TTS-based data augmentation and language model fusion, achieves a word error rate (WER) of \textbf{19.34\%} and a character error rate (CER) of \textbf{5.09\%} on the test set with a \textbf{57.02\% out-of-vocabulary (OOV) rate}. Furthermore, by filtering out minor errors involving missing or extraneous cedillas ($\cedilla$), the WER and CER further improve to \textbf{14.32\%} (26.48\% on unseen words) and \textbf{3.45\%}, respectively. These findings highlight the system’s capability to significantly expedite the transcription process for \sencoten, providing valuable assistance in efforts to revitalize the language.
\end{enumerate}

The rest of the paper is organized as follows. Section~\ref{sec:pipeline} outlines the proposed ASR-driven pipeline developed to support the documentation of the \sencoten language. Section~\ref{sec:aug} details the TTS system for synthesizing audio to augment ASR training data. Section~\ref{sec:cross} discusses the application of cross-lingual transfer learning based on speech foundation models. Experimental results and analysis on the \sencoten dataset are presented in Section~\ref{sec:exp}. Section~\ref{sec:conclusion} provides conclusions and discusses potential directions for future research.

\section{ASR-Driven Pipeline for \sencotensectitle Language Documentation}
\label{sec:pipeline}

As illustrated in Figure~\ref{fig:workflow}, our proposed ASR-driven pipeline for \sencoten language documentation consists of four stages, with carefully designed procedures to maximize the usage of the available audio and text data:

\noindent \textbf{Stage 1: Train the TTS system:} Parallel audio and text data in \sencoten are used to train a custom-designed text-to-speech (TTS) system, which will be described in detail in Section~\ref{sec:aug}.

\noindent \textbf{Stage 2: Generate synthesized audio via TTS:} \sencoten text without accompanying audio is fed into the trained TTS system to generate the corresponding synthesized audio.

\noindent \textbf{Stage 3: Perform cross-lingual transfer learning on the SFM:} The original parallel audio and text data, combined with the synthesized audio from the text-only data, are utilized to perform cross-lingual transfer learning on the speech foundation model (SFM), which will be outlined in Section~\ref{sec:cross}.

\noindent \textbf{Stage 4: Transcribe new audio with the fine-tuned SFM:} New audio in \sencoten is transcribed using the fine-tuned SFM, with the option to fuse an external language model (LM) trained on part or all of the available text data to further improve accuracy.

\section{Text-to-Speech Synthesis}
\label{sec:aug}

\begin{figure*}[!htbp]
    \centering
    \includegraphics[width=0.66\textwidth]{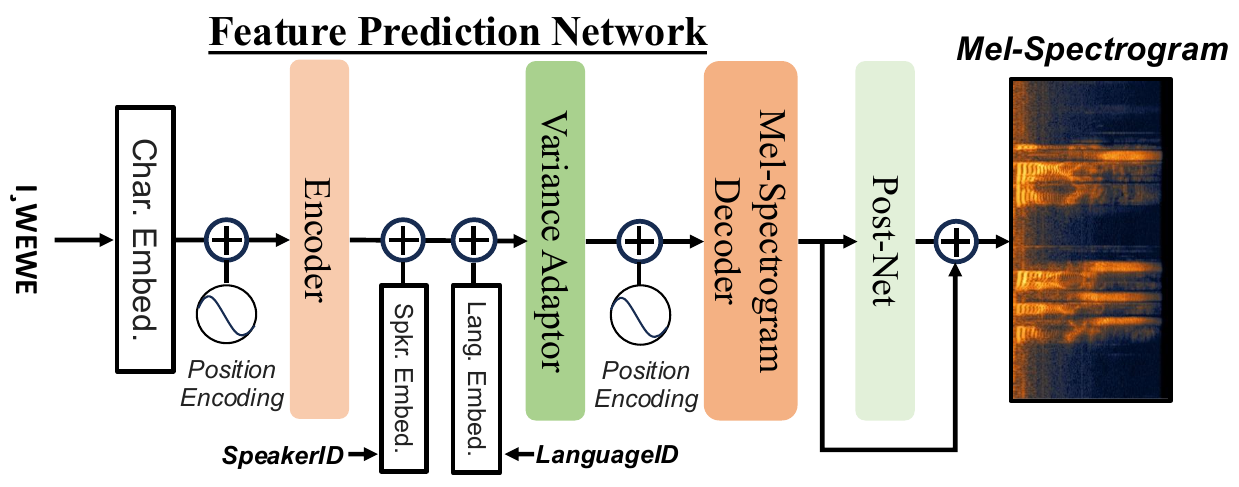}
    \caption{Architecture of the feature prediction work of our TTS system. ``Char. Embed.'', ``Spkr. Embed.'' and ``Lang. Embed.'' respectively denote character, speaker, and language embeddings.}
    \label{fig:feature_prediction}
\end{figure*}

Text-to-speech (TTS) synthesis has emerged as a powerful technique for augmenting ASR training datasets~\citep{gokay2019improving}, particularly in scenarios where parallel audio and text resources are limited.  As there exists written \sencoten text without corresponding audio recordings, TTS-generated audio can be used to augment the training data for developing \sencoten ASR systems.

To mitigate the scarcity of parallel audio \& text data in \sencoten, a three-phase approach is adopted using the EveryVoice TTS Toolkit~\citet{pine-etal-2022-requirements}, including \textbf{1)} training a feature prediction network, \textbf{2)} developing a vocoder, and \textbf{3)} aligning vocoder outputs with mel-spectrograms generated by the prediction network (i.e., vocoder matching).

\subsection{Feature Prediction Network}
\label{sec:feature_prediction}

Building on the work of~\citet{pine-etal-2022-requirements}, the EveryVoice TTS toolkit uses a modified FastSpeech 2~\citep{ren2020fastspeech} architecture as the feature prediction network. As illustrated in Figure~\ref{fig:feature_prediction}, the key modifications are:

\begin{itemize}
    \item Substitution of standard convolutions with depthwise separable convolutions in both the encoder and mel-spectrogram decoder~\citep{pine-etal-2022-requirements} to enhance parameter efficiency.
    \item Integration of learnable speaker embeddings (Figure~\ref{fig:feature_prediction}, middle, in circled box).
    \item Incorporation of a decoder post-net (Figure~\ref{fig:feature_prediction}, right, in light green).
\end{itemize}

Moreover, pre-generated forced alignments are replaced by a jointly-trained alignment module~\citep{badlanialignment}, while pitch and energy are predicted at the phoneme level instead of the frame level to achieve smoother prosody.

\subsection{Vocoder}
\label{sec:vocoder}
Since speech foundation models directly process raw audio, ensuring high-quality waveform synthesis is crucial. The vocoder, which converts intermediate mel-spectrograms into waveforms, plays a key role in this process. To this end, we utilize HiFi-GAN \citep{kong2020hifi}, a widely adopted generative adversarial network recognized for generating natural and high-quality waveforms, as the vocoder in our TTS system.

\subsection{Vocoder Matching}
\label{sec:vocoder_matching}
To mitigate the artifacts arising from limited training data, a vocoder matching strategy is employed after the initial training of the vocoder. This process fine-tunes the vocoder using mel-spectrograms generated by the feature prediction network as input, aligning it with the specific characteristics of these spectrograms to minimize discrepancies between training and inference conditions.

\section{Cross-Lingual Transfer Learning}
\label{sec:cross}

The limited availability of \sencoten data makes it impractical to train an end-to-end (E2E) ASR system from scratch. Alternatively, recent advances in speech foundation models (SFMs), which are pre-trained on large-scale datasets, offer a promising pathway for cross-lingual transfer learning in low-resource languages like \sencoten.

SFMs can be categorized into two main types:

\noindent \textbf{Encoder-Based SFM:}  
Encoder-based SFMs have gained widespread adoption due to their ability to convert raw audio into representations useful for various downstream tasks. Widely recognized models in this category include Wav2Vec 2.0 (Wav2Vec2)~\citep{baevski2020wav2vec}, HuBERT~\citep{hsu2021hubert}, WavLM~\citep{chen2022wavlm}, and Data2Vec~\citep{baevski2022data2vec}. These models employ a single encoder architecture to process audio, with a focus on self-supervised learning from unlabeled data. Both Wav2Vec2 and HuBERT excel at capturing rich speech representations, which are crucial for ASR in low-resource settings. WavLM further improves performance by effectively modeling not only speech but also environmental noise, making it particularly robust in challenging acoustic conditions. Data2Vec, on the other hand, expands the applicability of these models by generalizing the approach to multiple modalities.

\noindent \textbf{Encoder-Decoder-Based SFM:}  
In contrast, encoder-decoder-based SFMs integrate both an encoder to process the input audio and a decoder to generate transcriptions or other forms of output. Whisper~\citep{radford2023robust} is among the most well-known models in this category. By combining these two components, Whisper is capable of end-to-end transcription, making it a powerful tool for ASR tasks. Its architecture is particularly useful for languages with limited resources, as the encoder-decoder framework allows for more sophisticated handling of complex linguistic structures through cross-lingual transfer learning.

\section{Experiments}
\label{sec:exp}

\subsection{Task Description}
\label{sec:task}
\noindent \textbf{Parallel Audio \& Text Data:} The \sencoten speech dataset, part of the ``Speech Generation for Indigenous Language Education'' project~\citep{SGILE2025}, consists of about 4 hours of single-speaker recordings.  A Kaldi-based~\citep{povey2011kaldi} GMM-HMM system is used to estimate emission probabilities for each utterance. 20\% of the data is then allocated as the test set based on these estimates to ensure a balanced representation of difficulty. After silence stripping, the training set contains 1.7 hours and the test set 0.2 hours, with average utterance lengths of 2.06 and 2.04 seconds, respectively. The training set consists of 3k utterances with 3.6k distinct words, while the test set includes 0.8k utterances with 1.2k distinct words. The average word length is 3.6 characters in both sets. Due to \sencoten's polysynthetic nature and stress-driven metathesis, the test set shows a high out-of-vocabulary (OOV) rate of 57.02\%.

\noindent \textbf{Text-Only Data:} We have permission to access the \sencoten dictionary~\citep{montlerdictionary}, the most comprehensive lexicographic resource for the language, containing over 30k words and example sentences. This data is text-only, with no corresponding audio. 27k words and sentences are retained after filtering out overly long entries exceeding 81 characters.


\subsection{Experiment Setup}
\label{sec:setup}
\noindent{\textbf{Data processing:}}  We conduct silence stripping using SoX\footnote{\url{https://linux.die.net/man/1/sox}} and denoise the audio with an RNN-based denoiser\footnote{\url{https://github.com/xiph/rnnoise}}. The audio is then resampled to 16 kHz for ASR and 22.05 kHz for TTS development, while words are segmented into characters.

\noindent \textbf{Text-to-Speech Synthesis:} The TTS system outlined in Section~\ref{sec:aug} is built using the EveryVoice TTS Toolkit\footnote{\url{https://github.com/EveryVoiceTTS/EveryVoice}}. The train/test split mirrors that of the ASR system. The modified FastSpeech2 feature prediction network includes 4 encoder and 4 decoder blocks\footnote{Both encoder and decoder blocks have a 1024-dim feed-forward layer and two 128-dim attention heads.}, while HiFi-GAN in its V1 configuration~\citep{kong2020hifi} is used as the vocoder. The synthesized audio is automatically evaluated using the TorchSquim~\citep{torchsquim} model, which provides estimates for short-time objective intelligibility (STOI), perceptual evaluation of speech quality (PESQ), scale-invariant signal-to-noise ratio (SI-SNR), and mean opinion score (MOS).

\noindent \textbf{Cross-lingual Transfer Learning:} We utilize the Hugging Face platform to perform cross-lingual transfer learning with both encoder-based speech foundation models, including Wav2Vec2, HuBERT, WavLM, and Data2Vec, as well as encoder-decoder-based models\footnote{\url{https://huggingface.co/blog/{fine-tune-wav2vec2-english,fine-tune-whisper}}} like Whisper. SFMs of varying model sizes and pretraining data serve as the starting point for this process.

\noindent \textbf{Language Model Fusion:} We construct two 4-gram language models (LMs) using the KenLM toolkit~\citep{heafield2011kenlm}: 

\begin{itemize}
    \item A smaller model (``small-4g'') trained exclusively on the text from the training set.
    \item A larger model (``large-4g'') that also incorporates the 27k text-only \sencoten sentences.
\end{itemize}

The ``small-4g'' LM covers 3.6k words, while the ``large-4g'' spans 14k. For encoder-based SFMs (e.g., Wav2Vec2), shallow fusion~\citep{kannan2018analysis} integrates the n-gram LM during decoding. For encoder-decoder SFMs (e.g., Whisper), the LM rescales the n-best hypothesis list.

\subsection{Performance Analysis}
\label{sec:performance}

\begin{table}[htbp]
\caption{TTS evaluation on the \sencoten test set.}
\label{tab:tts}
\centering
    \scalebox{0.68}{\begin{tabular}{c|c|c|c|c}
    \toprule
        \textbf{Vocoder Matching} & \textbf{STOI} (\(\uparrow\)) & \textbf{PESQ} (\(\uparrow\)) & \textbf{SI-SNR} (\(\uparrow\)) & \textbf{MOS} (\(\uparrow\)) \\ 
        \midrule
        \xmark & 0.980 & 3.207 & 20.339 & 4.227 \\
        \cmark & 0.985 & 3.324 & 20.809 & 4.336 \\
    \bottomrule
    \end{tabular}
}
\end{table}

\begin{table*}[!htbp] 
\centering
    \caption{Performance of cross-lingual transfer learning using SFMs with different architectures.  \xmark~in the "Multilingual" column indicates that the SFM is pre-trained on English data only. "WER/CER" represents word/character error rate, while "seen" and "unseen" refer to whether the test words are included in the original training data.}
    \label{tab:ssl}
    \scalebox{0.85}{\begin{tabular}{c|c|c|c|c|cc|c}
    \toprule
        \multirow{2}{*}{\textbf{Sys.}} &
        \multirow{2}{*}{\textbf{Model}} &
        \multirow{2}{*}{\textbf{Multi-Lingual}} &
        \multirow{2}{*}{\textbf{LM}} &
        \multirow{2}{*}{\textbf{CER\% (\(\downarrow\))}} &
        \multicolumn{3}{c}{\textbf{WER\% (\(\downarrow\))}} \\ 
    \cmidrule(lr){6-8}
    & & & & & \textbf{seen} & \textbf{unseen} & \textbf{all} \\ 
    \midrule
        \textit{1} & \textit{Wav2Vec2-random-int} & \textit{-} & \textit{-} & \textit{84.36} & \textit{99.94} & \textit{100.00} & \textit{99.96} \\
    \midrule
        2 & Wav2Vec2-base & \multirow{2}{*}{\xmark} & \multirow{2}{*}{-} & 10.68 & 36.18 & 65.99 & 49.10 \\
        3 & Wav2Vec2-large & & & 8.25 & 21.42 & 56.87 & 35.04 \\
    \midrule
        4 & Wav2Vec2-xlsr-53 & \multirow{3}{*}{\cmark} & \multirow{3}{*}{-} & 11.23 & 33.48 & 69.13 & 51.44 \\
        5 & Wav2Vec2-xls-r-300m & & & 10.41 & 31.55 & 63.35 & 45.63 \\
        6 & Wav2Vec2-xls-r-1b & & & \textbf{6.32} & 14.87 & \textbf{55.04} & \textbf{27.81} \\
    \midrule
        7 & Data2Vec-base & \multirow{2}{*}{\xmark} & \multirow{2}{*}{-} & 14.89 & 40.09 & 78.56 & 60.61 \\
        8 & Data2Vec-large & & & 9.29  & 27.75 & 60.08 & 40.51 \\ 
    \midrule
        9 & HuBERT-base & \multirow{2}{*}{\xmark} & \multirow{2}{*}{-} & 13.34 & 45.15 & 72.04 & 58.61 \\
        10 & HuBERT-large & & & 12.29 & 44.66 & 69.78 & 56.06 \\
    \midrule
        11 & WavLM-base & \multirow{2}{*}{\xmark} & \multirow{2}{*}{-} & 11.70 & 36.18 & 71.27 & 50.71 \\
        12 & WavLM-large & & & 13.43 & 45.98 & 71.88 & 59.61 \\
    \midrule
        13 & Whisper-medium-en & \xmark & \multirow{2}{*}{-} & 7.36 & 15.38 & 57.30 & 28.13 \\
        14 & Whisper-large-v2 & \cmark & & \textbf{7.11} & 14.60 & \textbf{58.01} & \textbf{27.66} \\
    \midrule
        15 & \multirow{2}{*}{Wav2Vec2-xls-r-1b} & \multirow{2}{*}{\cmark} & small-4g & 6.05 & 12.18 & 57.92 & 25.13 \\
    \cmidrule(lr){4-8}
        16 & & & large-4g & \textbf{5.63} & 12.35 & \textbf{49.09} & 23.16 \\
    \midrule
        17 & \multirow{2}{*}{Whisper-large-v2} & \multirow{2}{*}{\cmark} & small-4g & 6.53 & 11.09 & 60.14 & 25.16 \\ 
    \cmidrule(lr){4-8}
        18 & & & large-4g & 6.12 & 10.95 & 50.35 & \textbf{22.67} \\
    \bottomrule
    \end{tabular}}
\end{table*}

\noindent The evaluation of the TTS system, cross-lingual transfer learning with SFMs, and the integration of TTS-based data augmentation and language models is conducted on the test set described in Section~\ref{sec:task}. In this context, ``seen'' and ``unseen'' words in terms of word error rate (WER) refer to whether the test words were present in the original training data.

\noindent \textbf{Text-to-Speech Synthesis:} We evaluate the TTS system on the \sencoten test set using four metrics: STOI, PESQ, SI-SNR, and MOS. As indicated in Table~\ref{tab:tts}, performance improves across all four metrics with vocoder matching. Based on this, we use the vocoder-matched system to synthesize 27k \sencoten sentences outlined in Section~\ref{sec:task}, resulting in approximately 11.6 hours of generated speech for ASR data augmentation.  Compared to the 13.3-hour augmented training set, the test set retains an OOV rate of 29.96\%.

\noindent \textbf{Cross-lingual Transfer Learning:} Table~\ref{tab:ssl} illustrates the results of cross-lingual transfer learning across various speech foundation models (SFMs). As part of an ablation study (Sys. 1), we also carry out an additional experiment where the weights of the Wav2Vec2-base model are randomly re-initialized to serve as the starting point. In addition, the top-performing encoder- and encoder-decoder-based SFMs are further integrated with the 4-gram LMs (Sys. 15-18).

Several insights can be drawn from Table~\ref{tab:ssl}: \textbf{1)} Larger SFMs do not consistently deliver better results than smaller models with similar architectures (Sys. 12 \textit{vs.} 11). \textbf{2)} Although the top-performing SFMs are pre-trained on multilingual datasets (Sys. 6, 14), they do not always outperform monolingual models with similar structures trained solely on English (Sys. 4-5 \textit{vs.} 3). \textbf{3)} Incorporating an external LM further boosts performance (Sys. 15-16 \textit{vs.} 6 and Sys. 17-18 \textit{vs.} 14), with larger LMs providing better outcomes (Sys. 16 \textit{vs.} 15, Sys. 18 \textit{vs.} 17). \textbf{4)} A substantial performance gap exists between words covered (``seen'') in the training data and those that are not (``unseen''), while the top-performing systems (Sys. 16,18) correctly transcribe roughly half of the unseen words.

\begin{table}[!htbp]
\centering
    \caption{Performance of incorporating TTS-synthesized data in cross-lingual transfer learning with the \textbf{Whisper} model. The original training data is always included, while ``all'' denotes the full 11.6-h synthesized data.}
    \label{tab:aug}
    \scalebox{0.77}{\begin{tabular}{c|c|c|c|cc|c}
        \toprule
            \multirow{2}{*}{\textbf{Sys.}} & 
            \multirow{2}{*}{\textbf{{\tabincell{c}{Aug.\\Data}}}} &
            \multirow{2}{*}{\textbf{LM}} & 
            \multirow{2}{*}{\textbf{CER\% (\(\downarrow\))}} &
            \multicolumn{3}{c}{\textbf{WER\% (\(\downarrow\))}} \\ 
        \cmidrule(lr){5-7}
            & & & & \textbf{seen} & \textbf{unseen} & \textbf{all} \\
        \midrule
            1 & 1h & \multirow{6}{*}{-} & 6.67 & 12.97 & 54.97 & 25.38 \\
            2 & 2h &  & 6.02 & 12.67 & 51.42 & 23.99 \\
            3 & 4h &  & 5.84 & 12.67 & 47.52 & 22.86 \\
            4 & 6h &  & 5.72 & 11.34 & 49.79 & 22.56 \\
            5 & 8h &  & 5.79 & 11.79 & 48.44 & 22.53 \\
            6 & all &  & \textbf{5.63} & 12.18 & \textbf{44.81} & \textbf{22.01} \\
        \midrule
            7 & 1h & \multirow{6}{*}{{\tabincell{c}{small\\fg}}} & 6.80 & 10.55 & 56.31 & 23.74 \\
            8 & 2h &  & 5.82 & 10.95 & 52.41 & 22.82 \\
            9 & 4h &  & 5.50 & 9.71 & 49.79 & 21.21 \\
            10 & 6h &  & 5.59 & 9.96 & 50.50 & 21.65 \\
            11 & 8h &  & 5.42 & 9.57 & 48.73 & 20.70 \\
            12 & all &  & \textbf{5.26} & 9.91 & \textbf{46.51} & \textbf{20.51} \\
        \midrule
            13 & 1h & \multirow{6}{*}{{\tabincell{c}{large\\fg}}}& 6.60 & 10.65 & 49.08 & 22.60 \\
            14 & 2h &  & 5.96 & 10.06 & 49.36 & 22.42 \\
            15 & 4h &  & 5.38 & 9.57 & 45.67 & 20.84 \\
            16 & 6h &  & 5.18 & 9.22 & 46.60 & 20.44 \\
            17 & 8h &  & \textbf{5.09} & 8.43 & 44.96 & \textbf{19.34} \\
            18 & all &  & 5.11 & 9.62 & \textbf{43.53} & 20.15 \\
        \bottomrule
    \end{tabular}}
\end{table}

\noindent \textbf{TTS-Based Data Augmentation:} We progressively incorporate TTS-synthesized data into the cross-lingual transfer learning process of the top-performing SFM (Table~\ref{tab:ssl}, Sys. 18), i.e., Whisper\footnote{\url{https://huggingface.co/openai/whisper-large-v2}}. The augmentation begins with 1 hour of synthesized data and scales up to a total of 11.6 hours. As shown in Table~\ref{tab:aug}, several trends can be observed: \textbf{1)} Incorporating TTS-synthesized data leads to ASR performance improvements both with or without an external LM (Sys. 1-6, 7-12, 13-18  in Table~\ref{tab:aug} \textit{vs.}  Sys. 14,17,18 in Table~\ref{tab:ssl}), with overall WER reductions of up to 5.65\% abs. (20.43\% rel.) and 13.20\% abs. (22.75\% rel.) on unseen words absent from the original training set (Sys. 6 in Table~\ref{tab:aug} \textit{vs.} Sys. 14 in Table~\ref{tab:ssl}). \textbf{2)} For systems fused with the large 4-gram LM covering all text used for TTS, the inclusion of synthesized audio further improves ASR performance (Sys. 13-18 in Table~\ref{tab:aug} \textit{vs.} Sys. 18 in Table~\ref{tab:ssl}). \textbf{3)} There is a general trend of performance convergence when 8 hours of TTS-synthesized data are added (Sys. 5,11,17 in Table~\ref{tab:aug}).

\subsection{Language Documentation Support}

The motivation for this project stemmed from discussions between the authors in the context of regular meetings related to a multi-year TTS research project, described in detail in \citet{SGILE2025}. ASR was not an explicit goal of the project that brought us together, but the first author of this paper has expertise in speech recognition and realized that we had some of the requisite pieces to develop a proof-of-concept ASR system, namely, an established relationship with the language community in question, pre-trained TTS models, and some modest amounts of parallel text-audio data. The first author proposed the idea, along with possible benefits and risks, to members of the \wsanec school Board at a meeting for the TTS project, which was met with enthusiasm and support leading to this initial effort. Despite the strong results from these initial experiments, many more steps and protocols will be required to connect this technology with on-the-ground language efforts.

\begin{figure*}[!t]
    \centering
    \includegraphics[width=0.75\textwidth]{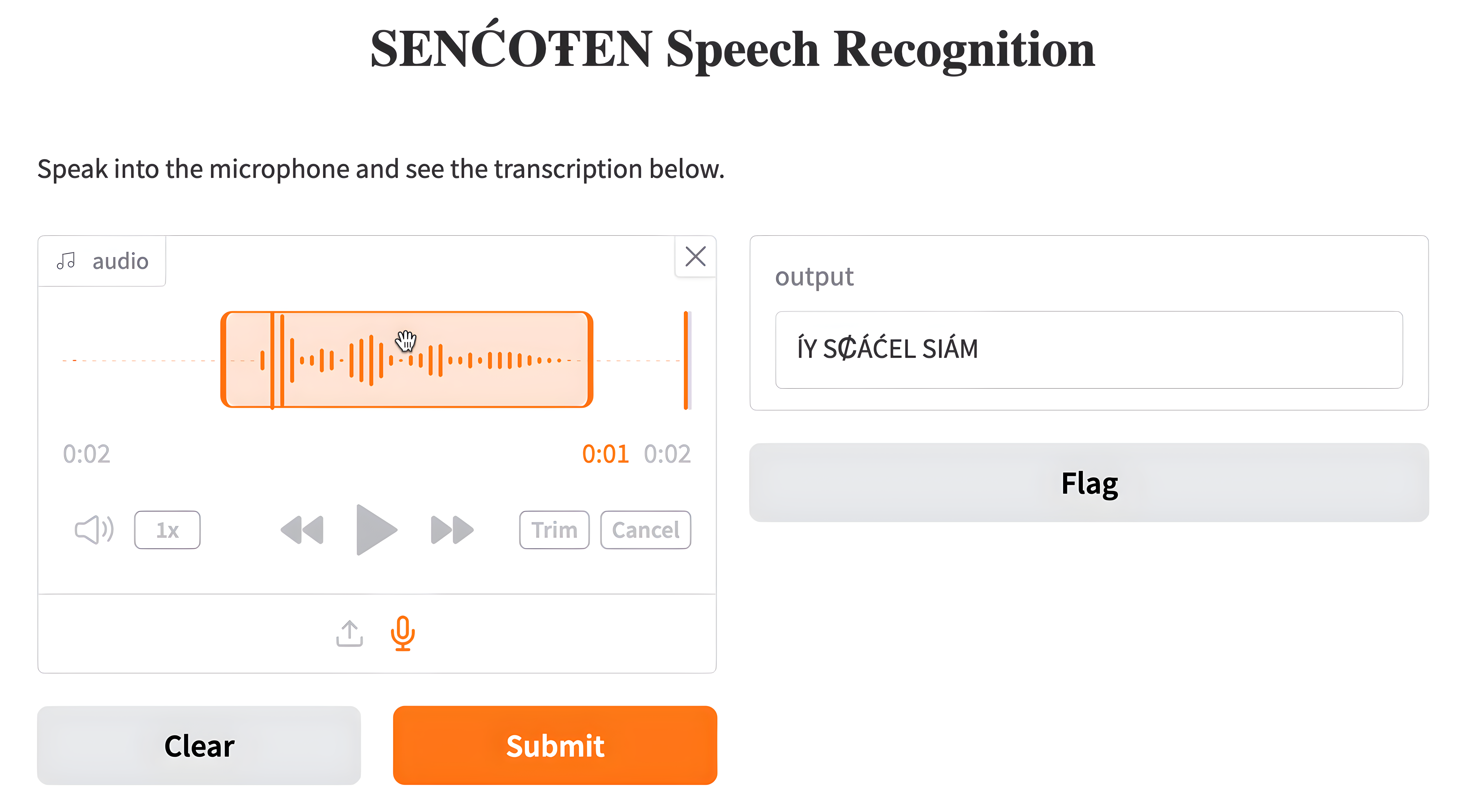}
    \caption{Demonstration of the user interface designed to support \sencoten language documentation.}
    \label{fig:demo}
\end{figure*}

To help us demonstrate the capabilities of this technology, we developed an intuitive, web-based user interface and API using the Gradio framework~\citep{abid2019gradio} in Python. The interface, as illustrated in Figure~\ref{fig:demo}, allows users to interact with the model by either speaking directly into the microphone or uploading a pre-recorded audio file. After providing the input, users can click the ``Submit'' button (Figure~\ref{fig:demo}, bottom, in orange) to generate an automatic transcription, displayed in the text box labeled ``output'' (Figure~\ref{fig:demo}, right).



Additionally, users can select specific segments of the audio (Figure~\ref{fig:demo}, left) to view their corresponding transcriptions, enabling precise analysis of smaller portions of the recording. The ``Flag'' button, located on the right, allows users to mark an audio-transcription pair for further review, annotation, or reference. This functionality is particularly useful for collaborative workflows, where flagged segments may require validation or additional context from language experts. By simplifying transcription workflows, the interface streamlines language documentation, enabling linguists, community members, and researchers to efficiently process spoken language data with decreased manual effort. Features like audio segmentation and flagging support iterative transcription processes, while the web-based design ensures accessibility across a wide range of devices, making it well-suited for teams working in different locations.

To safeguard the model's privacy, the interface is currently accessible exclusively through a private web gateway. Future developments aim to facilitate closer alignment with documentation and language revitalization workflows \citep[e.g.,][]{cox_perspehone2019,adams-etal-2021-user}.

A closer examination of the decoded outputs from the \sencoten ASR systems reveals that a notable portion of the errors involve missing or extraneous cedillas ($\cedilla$) which indicate either glottalization when following resonant or glottal stops otherwise. Given that these errors are relatively easily correctable by a \sencoten speaker, and that the consistency of their use varies, we reassess the ASR performance with these errors excluded. As shown in Table~\ref{tab:remove_error}, the system achieves an overall WER of \textbf{14.32\%}, a CER of \textbf{3.45\%}, and a WER of \textbf{26.48\%} on unseen words. This demonstrates the potential of our proposed ASR-driven pipeline to support the documentation for \sencoten.

\begin{table}[ht]
    \centering
    \caption{Performance of the top-performing SFM (Sys. 17 in Table~\ref{tab:aug}), excluding errors related to cedilla ($\cedilla$).}
    \label{tab:remove_error}
    \scalebox{0.85}{\begin{tabular}{c|c|cc|c}
        \toprule
            \multirow{2}{*}{\textbf{Model}} &
            \multirow{2}{*}{\textbf{CER\% (\(\downarrow\))}} &
            \multicolumn{3}{c}{\textbf{WER\% (\(\downarrow\))}} \\
        \cmidrule(lr){3-5}
            & & \textbf{seen} & \textbf{unseen} & \textbf{all} \\ 
        \midrule
            Whisper & \textbf{3.45} & 6.88 & \textbf{26.48} & \textbf{14.32} \\ 
        \bottomrule
    \end{tabular}}
\end{table}

\section{Conclusion}
\label{sec:conclusion}
In this paper, we proposed an ASR-driven pipeline designed to tackle the unique challenges of documenting the \sencoten language, which is hindered by data scarcity, substantial vocabulary variation, and phonological complexity. By incorporating augmented speech data from a TTS system, cross-lingual transfer learning using speech foundation models (SFMs), and an n-gram language model via shallow fusion, we demonstrated the effectiveness of our approach in improving ASR performance for low-resource languages. Our experiments on the \sencoten dataset yielded a WER of 19.34\% and a CER of 5.09\%, with further improvements to a WER of 14.32\% (26.48\% on unseen words) and a CER of 3.45\% after mitigating minor cedilla-related errors. These results highlight the potential of the proposed pipeline to enhance \sencoten language documentation, offering a valuable tool for ongoing language revitalization efforts. Future work will focus on more linguistically oriented techniques, for example, modeling stress-driven metathesis in \sencoten.



\bibliography{custom}

\begin{appendices}

\section{Further Ablation Studies}
\label{sec:appendix_ablation}

To get an in-depth analysis of our proposed ASR-driven documentation pipeline (Figure~\ref{fig:workflow}), two sets of ablation studies are further conducted: \textbf{1)} replacing the end-to-end speech foundation model with a conventional hybrid TDNN ASR system, and \textbf{2)} Reducing the training data to as little as 10 mins to simulate ultra-low resource settings.

\noindent \textbf{Hybrid TDNN Systems:} The hybrid TDNN system is constructed following the Kaldi Chain recipe\footnote{\url{https://github.com/kaldi-asr/kaldi/tree/master/egs/librispeech/s5}} but with a more compact architecture featuring 7 context-splicing layers with time strides of $\{1,1,0,3,3,6\}$. I-Vectors~\citep{saon2013speaker} are incorporated while speed perturbation is omitted. The text is transcribed into International Phonetic Alphabet (IPA) representations using the g2p library~\cite{pine2022gi2pi}. 

Table~\ref{tab:tdnn} reveals the following trends: \textbf{1)} Using a larger language model (LM) leads to noticeable performance degradation when the additional words in the LM lack corresponding audio in the training set (Sys. 2 \textit{vs.} Sys. 1). \textbf{2)} Using the small 4-gram, expanding the training set's word coverage with TTS-synthesized data leads to marginal performance improvement (Sys. 3 \textit{vs.} Sys. 1). However, a substantial gain is achieved when the text used for TTS is also included in LM training (Sys. 4 \textit{vs.} Sys. 3). \textbf{3)} The top-performing SFMs (Sys. 17-18 in Table~\ref{tab:aug}) largely outperforms the best hybrid TDNN system (Sys. 4 in Table~\ref{tab:tdnn} across all metrics, showing the effectiveness of using SFMs in the proposed ASR-driven documentation pipeline.


\begin{table}[htbp]
\caption{Performance of hybrid TDNN system. ``Data Aug.'' refers to TTS-based data augmentation. ``\# Hrs'' denotes the duration of the training set.}
\label{tab:tdnn}
\centering
    \scalebox{0.66}{\begin{tabular}{c|c|c|c|c|cc|c}
    \toprule
        \multirow{2}{*}{\textbf{Sys.}} &
        \multirow{2}{*}{\tabincell{c}{\textbf{Data}\\\textbf{Aug.}}} &
        \multirow{2}{*}{\textbf{\# Hrs}} &
        \multirow{2}{*}{\textbf{LM}} &
        \multirow{2}{*}{\textbf{CER\% (\(\downarrow\))}} &
        \multicolumn{3}{c}{\textbf{WER\% (\(\downarrow\))}} \\
    \cmidrule(lr){6-8}
        & & & & & \textbf{seen} & \textbf{unseen} & \textbf{all} \\ 
    \midrule
        1 &
        \multirow{2}{*}{\xmark} &
        \multirow{2}{*}{1.7} &
        small-4g & 19.68 & 18.93 & 100.00 & 46.92 \\ 
    \cmidrule(lr){1-1}\cmidrule(lr){4-8}
        2 & & & large-4g & 19.59 & 20.46 & 100.00 & 49.34 \\ 
    \cmidrule(lr){1-1}\cmidrule(lr){2-8}
        3 & \multirow{2}{*}{\cmark} & \multirow{2}{*}{13.3} &
        small-4g & 19.72 & 16.67 & 100.00 & 46.23 \\
    \cmidrule(lr){1-1}\cmidrule(lr){4-8}
        4 & & & large-4g & \textbf{9.65} & 16.62 & \textbf{58.92} & \textbf{36.63} \\ 
    \bottomrule
    \end{tabular}
    }
\end{table}

\noindent \textbf{Ultra-Low Resource Scenarios:} We simulate ultra-low-resource conditions by utilizing just 10 minutes of parallel audio and text data to perform cross-lingual transfer learning with SFMs, excluding the external LM, and assuming this limited training data is the only available resource.  As shown in Table~\ref{tab:limit_data}, Whisper (Sys. 5-8) is more sensitive to the amount of data available for transfer learning compared to Wav2Vec2 (Sys. 1-4). This may be attributed to differences in their architectures, with Wav2Vec2 being encoder-based, while Whisper follows an encoder-decoder structure.

\begin{table}[!htbp]
\centering
    \caption{Performance of cross-lingual transfer learning on SFMs in ultra-low resource scenarios with as little as 10 min training data. No LM fusion is incorporated.}
    \label{tab:limit_data}
    \scalebox{0.7}{\begin{tabular}{c|c|c|c|cc|c}
    \toprule
        \multirow{2}{*}{\textbf{Sys.}} &
        \multirow{2}{*}{\textbf{Model}} &
        \multirow{2}{*}{\tabincell{c}{\textbf{Train}\\\textbf{Data}}} &
        \multirow{2}{*}{\textbf{CER\% (\(\downarrow\))}} &
        \multicolumn{3}{c}{\textbf{WER\% (\(\downarrow\))}} \\
    \cmidrule(lr){5-7}
        & & & & \textbf{seen} & \textbf{unseen} & \textbf{all} \\
    \midrule
        1 & \multirow{4}{*}{Wav2Vec2} & all & 6.32 & 14.87 & 55.04 & 27.81 \\
        2 & & 1h & 7.54 & 20.15 & 54.99 & 32.92 \\
        3 & & 30min & 8.78 & 23.79 & 59.77 & 37.74 \\
        4 & & \textbf{10min} & \textbf{12.11} & 34.42 & \textbf{67.94} & \textbf{50.13} \\
    \midrule
        5 & \multirow{4}{*}{Whisper} & all & 7.11 & 14.60 & 58.01 & 27.66 \\
        6 & & 1h & 9.13 & 17.36 & 59.94 & 31.17 \\
        7 & & 30min & 10.95 & 26.53 & 69.22 & 41.17 \\
        8 & & 10min & 21.28 & 43.34 & 82.58 & 63.00 \\
    \bottomrule
    \end{tabular}}
\end{table}
\end{appendices}
\end{document}